\documentclass[10pt]{article}
\usepackage[utf8]{luainputenc}
\usepackage{textcomp}
\usepackage{amsmath}
\usepackage{amssymb}

\makeatletter

\newcommand{\lyxmathsym}[1]{\ifmmode\begingroup\def\b@ld{bold}
  \text{\ifx\math@version\b@ld\bfseries\fi#1}\endgroup\else#1\fi}

\makeatother

\begin{document}

\title{From Minkowskian Solution of General Relativity with Cosmological
Constant to the Accelerating Universe}

\author{Dr. Yves Pierseaux (ypiersea@ulb.ac.be)}
\maketitle
\begin{abstract}
A Minkowskian solution of the equation of General Relativity (as written
by Einstein in 1915) is trivial because it simply means that both
members of the equation are equal to zero. However, if alternatively,
one considers the complete equation with a non-zero constant $\Lambda$
(Einstein 1917), a Minkowskian solution is no longer trivial because
it amounts to impose a constraint on the right hand side of the equation
(i.e. a non-null stress-energy tensor). If furthermore one identifies
(as usual) this tensor to the one of a perfect fluid, one finds that
this fluid has a positive energy density and a negative pressure that
depend on the three constants of the equation (i.e. gravitational
constant $G$, cosmological constant $\Lambda$ and velocity of light
c). When doing that (§1), one has to consider the ''Minkowskian Vacuum''
as a physical object of $GR$ (an enigmatic non-baryonic Minkowskian
fluid).

Can one build a model of this object on the basis of a dynamical equilibrium
between the effective gravitational attraction due to the positive
energy density versus the negative pressure repulsion? We propose
to study such a model, where the (enigmatic) fluid is assumed to exist
only in a limited sphere whose surface acts like a ''test body''
sensitive to the gravitational field created by the fluid. No static
equilibrium exists, but a pseudoNewtonian ''dynamical equilibrium''
§2) can be reached if the pseudoEuclidean fluid is in state of expansion.
Up to there, we have simply constructed a model of an ''abstract
Universe'' (i.e. the limited sphere: There is no fluid outside this
sphere !) that gives to a (purely mathematical) constant $\Lambda$
a concrete physical meaning. 

We discover finally that our expanding fluid has not only dynamical
properties (§3) but also optical properties that are connected with
Doppler Redshift (§4). Remembering that recent observations in Cosmology
indicate that the ''real Universe'' seems to be ''Flat'' and in
''Accelerated Expansion''; remembering also (after all..) that the
archetypal Flat Universe is simply Minkowskian Universe, we logically
wonder if the unexpected Minkowskian global solution, could not be
also a significant cosmological model (conclusion). 
\end{abstract}

\section{\protect\bigskip{}
Enigmatic Minkowskian\ Fluid deduced from Complete Einstein's Equation }

Let us consider Einstein's basic equation (\cite{2}) of General Relativity
($GR)$ \textit{completed} by a positive mathematical constant $\Lambda>0$
( \cite{1}), that has a priori nothing to do with Cosmology (with
$g^{\mu\nu}$ Riemanian metric, $G^{\mu\nu}$ Einstein's curvature
tensor, $T^{\mu\nu}$ stress-energy tensor, $G$ gravitational constant
and $c$ light velocity):

\begin{equation}
G_{\mu\nu}+\Lambda g_{_{\mu\nu}}=\frac{8\pi G}{c^{4}}T_{\mu\nu}\label{1}
\end{equation}
In order to discover the physical meaning of this constant$\Lambda>0$,
let us simplify with $G_{\mu\nu}=0$ the equation (1) by introducing
Minkowskian metric $g_{_{\mu\nu}}=\eta_{\mu\nu}=\left(\begin{array}{c}
1\\
0\\
0\\
0
\end{array}\begin{array}{c}
0\\
-1\\
0\\
0
\end{array}\begin{array}{c}
0\\
0\\
-1\\
0
\end{array}\begin{array}{c}
0\\
0\\
0\\
-1
\end{array}\right)$. We obtain a tensor of Minkowskian Vacuum (2):

\begin{equation}
T_{\mu\nu}^{VACUUM}=\frac{\Lambda c^{4}}{8\pi G}\mathbf{\eta}_{\mu\nu}\label{2}
\end{equation}
Let us now associate to this tensor (2) the one of a perfect relativistic
fluid:

\begin{equation}
T_{\mu\nu}^{VACUUM}\equiv(p+\rho)\frac{u_{\mu}u_{\nu}}{c^{2}}-p\eta_{\mu\nu}\label{3}
\end{equation}
Minkowskian Vacuum is then simulated by an enigmatic fluid with a
positive density of energy $\rho=\frac{\Lambda c^{4}}{8\pi G}$ and
a negative pressure%
\footnote{This negative pressure of vacuum seems to be a stranger in $SR$.
However, on the only basis of $LT$ in 1905, Poincaré (1853-1912)
introduced such a negative pressure of ether in order to define a
punctual electron by its electromagnetic field (\cite{3}). He suspected
the gravitational origin of his negative pressure but he did not succeed
to find the exact relation between his ''gravitational'' scalar
field $p$ and the gravitational constant $G$.%
}: $p=-\frac{\Lambda c^{4}}{8\pi G}$ 
\begin{equation}
T_{\mu\nu}^{VACUUM}=T_{\mu\nu}^{GR}=\left(\begin{array}{c}
\rho\\
0\\
0\\
0
\end{array}\begin{array}{c}
0\\
p\\
0\\
0
\end{array}\begin{array}{c}
0\\
0\\
p\\
0
\end{array}\begin{array}{c}
0\\
0\\
0\\
p
\end{array}\right)=\frac{\Lambda c^{4}}{8\pi G}\left(\begin{array}{c}
1\\
0\\
0\\
0
\end{array}\begin{array}{c}
0\\
-1\\
0\\
0
\end{array}\begin{array}{c}
0\\
0\\
-1\\
0
\end{array}\begin{array}{c}
0\\
0\\
0\\
-1
\end{array}\right)\text{ \ \ \ \ \ }p+\rho=0\label{4}
\end{equation}
Our enigmatic Minkowskian fluid becomes a physical object in the framework
of (complete) $GR$. Before the examination of the physical properties
of our fluid determined by three basic constant ($\Lambda,$G and
$c$) (§2), let us formulate two remarks.

\textbf{\textit{REMARK 1 Our non-usual fluid (4) cannot be confused
with usual perfect fluid (4-SR) in the framework of standard Special
Relativity (}}\textit{$SR$}\textbf{\textit{)}}\textit{. In this case,
we have $u_{\mu}u_{\nu}=0$ in proper system for all components except
for purely temporal components $u_{0}u_{0}=c^{2}=1)$$ $
\begin{equation}
T_{\mu\nu}^{SR}=\left(\begin{array}{c}
p+\rho\\
0\\
0\\
0
\end{array}\begin{array}{c}
0\\
0\\
0\\
0
\end{array}\begin{array}{c}
0\\
0\\
0\\
0
\end{array}\begin{array}{c}
0\\
0\\
0\\
0
\end{array}\right)-\left(\begin{array}{c}
p\\
0\\
0\\
0
\end{array}\begin{array}{c}
0\\
-p\\
0\\
0
\end{array}\begin{array}{c}
0\\
0\\
-p\\
0
\end{array}\begin{array}{c}
0\\
0\\
0\\
-p
\end{array}\right)=\left(\begin{array}{c}
\rho\\
0\\
0\\
0
\end{array}\begin{array}{c}
0\\
p\\
0\\
0
\end{array}\begin{array}{c}
0\\
0\\
p\\
0
\end{array}\begin{array}{c}
0\\
0\\
0\\
p
\end{array}\right)\text{\ \ \ \ \ \ \ }p+\rho\neq0\tag{4\_SR}
\end{equation}
Any relativistic usual perfect fluid has a positive pressure }$p$\textit{\ (}$0\leq p\leq\frac{1}{3}w_{em}$\textit{)}%
\footnote{For example we have free electromagnetic (em) isotropic field$\left(w_{em},\text{ }p_{em},\text{ }p_{em},\text{ }p_{em}\right)$
with null trace $p_{em}=+\frac{1}{3}w_{em}$. According to Poincaré,
we have to add a special tensor to the tensor of em field in order
to define Poincaré's electron: ($c=1$): $T_{\mu\nu}^{electron}=$$T_{\mu\nu}^{em}+p_{em}\eta_{\mu\nu}=(p_{em}+w_{em})u_{\mu}u_{\nu}=\frac{4}{3}w_{em}u_{\mu}u_{\nu},$
In proper system, we have: $T_{\mu\nu}^{electron}=\left(\text{ }w_{em},p_{em},p_{em},p_{em}\right)+\left(\text{ }\frac{1}{3}w_{em},-p_{em},-p_{em},-p_{em}\right)=\left(\frac{4}{3}w_{em},0,0,0\right).$
Poincaré's special tensor (non-null trace) is then $\left(\begin{array}{c}
\frac{1}{3}w_{em}\\
0\\
0\\
0
\end{array}\begin{array}{c}
0\\
p_{P}\\
0\\
0
\end{array}\begin{array}{c}
0\\
0\\
p_{P}\\
0
\end{array}\begin{array}{c}
0\\
0\\
0\\
p_{P}
\end{array}\right)$corresponding to negative Poincaré's non-em pressure $p_{Poincar\acute{e}}=p_{P}=-\frac{1}{3}w_{em}$.
This is the reason why, one admits usually (von Laue) that $3/4$
of electron inertial mass ($\frac{4}{3}w_{em}=m_{e}$) comes from
electromagnetic origin. We showed in ``La Structure Fine de la Relativité
Restreinte'' that there is an enigmatic connection between Poincaré's
gravitational electron (1905) and Einstein's (gravitational) photon
(light complex or photon, 1905, see note 10).%
}\textit{. In this way our enigmatic fluid is no longer a usual fluid
in ''immutable Minkowskian Vacuum'' (standard }$SR$\textit{) but
it is the (Classical) Minkowskian Continuum itself . }

\textbf{\bigskip{}
}\textbf{\textit{REMARK 2 Our non-usual (classical) fluid (2) cannot
be confused with usual (quantum) black energy (2bis) in the framework
of Cosmology.}}\textit{ Standard method in Cosmology consists in associating
a supplementary stress-tensor $T_{\mu\nu}^{\Lambda}$ to a cosmological
constant (CC $\Lambda$) in the second member of (1) in order to have
a second contribution to $G_{\mu\nu}:$(with c=1):
\begin{equation}
G_{\mu\nu}=8\pi GT_{\mu\nu}-\Lambda g_{\mu\nu}=8\pi G(T_{\mu\nu}-T_{\mu\nu}^{\Lambda})\ \ \ \text{with}\ \ \ \ \ T_{\mu\nu}^{\Lambda}=\frac{\Lambda}{8\pi G}g_{\mu\nu}=\rho_{\Lambda}g_{\mu\nu}\tag{2\_Riemann}
\end{equation}
By associating }$T^{\mu\nu}$\textit{\ stress-energy tensor to the
one of a perfect relativistic fluid }$T_{\mu\nu}^{\Lambda}=(p+\rho)u_{\mu}u_{\nu}-pg_{\mu\nu}$\textit{we
obtain usually a fluid (black energy of ''quantum vacuum''}%
\footnote{Let us note that the ''vacuum''\ $T_{\mu\nu}^{\Lambda}$ in (2-Riemann)
depends on $g_{\mu\nu}$ ($G_{\mu\nu}\neq0$) that indicates a presence
of usual matter whilst the vacuum in (2) is radically without any
usual (baryonic) matter.%
}\textit{) characterized by an unknown Riemanian metric }$g_{\mu\nu}$\textit{\ (2-Riemann,
see 18-RW) whilst in (2) the metric is determined a priori Minkowskian.
In standard Cosmology, Minkowskian limit can only be a trivial result
of a very improbable compensation }$T_{\mu\nu}-T_{\mu\nu}^{\Lambda}=0.$

\section{Thermodynamical Properties of Minkowskian Fluid and Unstable Static
Model}

Basic condition $\rho+p=0$ (4) gives a new physical interpretation
of Minkowskian metric as a fluid. 
\begin{equation}
g_{_{\mu\nu}}=\eta_{\mu\nu}\qquad\Leftrightarrow\qquad\rho+p=0=h\tag{4-bis}
\end{equation}
The geodesic of a material point is usually determined in Minkowskian
space-time as a straight line. But here we have a point of space-time
continuum itself. In order to discover physical properties of our
enigmatic fluid the only possible point of departure is local thermodynamical
properties given by (4bis):where $h$ is \textit{null density of enthalpy.}
Given that $\rho$ is a density of energy of fluid, we have by integration
a finite volume\ $V$ with a finite energy $U$: 
\begin{equation}
U+pV=0=H\label{5}
\end{equation}
 By differentiation we obtain: 
\begin{equation}
dU+pdV=0=dH=h\label{6}
\end{equation}
that seems trivially return to (4bis) with reduction of element of
volume $dV\neq0$ ($dH=hdV=0$). Usually it is claimed that Minkowskian
vacuum would be static ($dV=0)$. Let us consider, at flat Minkowskian
limit, an Euclidean sphere of fluid:

\begin{equation}
V=\frac{4}{3}\pi R^{3}\ \ \ \ \ \ U=\frac{4}{3}\pi\rho R^{3}\label{7}
\end{equation}
 At Minkowskian limit we have also to take into account Einstein's
relation of ''materialization'' of energy:
\begin{equation}
U=Mc^{2}\qquad M=\frac{4}{3}\pi\frac{\rho}{c^{2}}R^{3}\label{8}
\end{equation}
How can we test the behavior (static or not static) of such an Euclidean
Sphere of fluid? Let us consider a \textit{test point} (infinitesimal
pseudomass%
\footnote{Like in electrostatic we consider a test charge (mass) as small as
possible in such a way that we have no modification of the electrostatic
field.%
} $\mu=dM$) on the surface of sphere. We have to introduce gravitational
constant because $\rho_{\text{VACUUM}}=\frac{\Lambda c^{2}}{8\pi G}$.
We suggest then to study a Newtonian model where the Minkowskian fluid
is assumed to exist only in a limited sphere whose surface acts like
a ''test body'' sensitive to the gravitational field created by
the fluid. The surface is submitted to gravitational attractive potential
\begin{equation}
\Phi=-\frac{GM}{R}\text{=\ensuremath{-}\ensuremath{\frac{4}{3}\pi G\frac{\rho}{c^{2}}R^{2}}\ \ \ \ \ \ \ \ \ \ \ }U_{P}=-\frac{GM\mu}{R}=-\ensuremath{\frac{4}{3}\pi G\frac{\rho}{c^{2}}R^{2}\mu}\label{9}
\end{equation}
Then the surface of fluid will collapse towards the center of sphere
given that we have only attractive potential energy. So a \textit{static}
finite sphere of our fluid is unstable $(\Delta V=V\longmapsto0)$
and Minkowskian solution (4bis) seems impossible. We rediscover in
this way that standard immutable Minkowskian vacuum must be defined
without gravitation.

\section{Dynamical Properties of fluid, Radial Expanding Universe and Scalar
Field of Gravitation}

The existence of our fluid is directly connected with Minkowskian
(Pseudo-Euclidean) space-time, where basically the time is not separated
from space (2). Let us thus consider that thermodynamical differential
$dV$ variation of volume of fluid is a \textit{temporal} variation
$dV(t)$: 
\begin{equation}
dU(t)+pdV(t)=0\label{10}
\end{equation}
In this way, equation (6) is no longer trivial. We have a variable
volume $V(t)$ coupled with a constant density 
\begin{equation}
U(t)=\rho V(t)=\frac{4}{3}\pi\rho R^{3}(t)=M(t)c^{2}\label{11}
\end{equation}
Let us now consider that (prerelativistic) Newtonian law of gravitation
is also variable with a temporal gravitational potential $\Phi(t)=-\frac{GM(t)}{R(t)}$.
Our test point $(\mu=dM)$ at radial distance $R(t)$ has therefore
a positive radial velocity $\frac{dR(t)}{dt}=$ $\dot{R}(t)$. Potential
energy $U_{P}=-\frac{G\mu M(t)}{R(t)}$ can be then now compensated
by kinetics energy $U_{C}=\frac{1}{2}\mu\dot{R}(t)^{2}$: 
\begin{equation}
\frac{1}{2}\dot{R}(t)^{2}-\frac{GM(t)}{R(t)}=0\qquad\Longrightarrow\qquad\frac{1}{2}\dot{R}(t)^{2}-\frac{4}{3}\pi G\frac{\rho}{c^{2}}R(t)^{2}=0\label{12}
\end{equation}
($\mu$ disappears). This Pseudo-Newtonian model%
\footnote{``Pseudo'' because Newton's law depending on time is not the usual
point of view. We adopt in this way, in accordance with a GR point
of view, that we can extend Newton's law of gravitation from material
baryonic particles m to points of space M. %
} of Pseudo-Euclidean fluid is based on a \textit{dynamical} equilibrium
``sphere-test body'' between attraction and repulsion. We obtain
in this way a stability of expanding sphere with a radial enigmatic
(Remark 3) ''escape velocity'' $\dot{R}(t)>0$: 
\begin{equation}
\dot{R}(t)=\frac{R(t)}{c}\sqrt{\frac{8}{3}\pi G\rho}=cR(t)\sqrt{\frac{\Lambda}{3}}\label{13}
\end{equation}
If we suppose a finite spherical volume of fluid in dynamical equilibrium
then it is in exponential expanding (11). Escape velocity (13) disappears
if and only if $\Lambda=0.$ Physical meaning of mathematical constant
$\Lambda$ is now clarified by Minkowskian solution that implies the
introduction (from 13) of a GLOBAL\ SCALE FACTOR $R(t)$ (with Minkowskian
metric, 2 or see 19): 
\begin{equation}
R(t)=R_{H}e^{\sqrt{\frac{\Lambda}{3}}t}\label{14}
\end{equation}
with a constant of integration $R(0)=R_{H}$ that seems, at first
sight, not depend on $\Lambda$. 

We can \ also define a constant of expansion of Fluid (Vacuum) that
we suggest to note $H_{\Lambda}$(15 left): 
\begin{equation}
H_{\Lambda}=\frac{\dot{R}(t)}{R(t)}\qquad\ \ \text{\ }\rho_{\text{VACUUM}}=\frac{3H_{\Lambda}^{2}c^{2}}{8\pi G}\label{15}
\end{equation}
together with a density \textit{inside} the sphere (15 right). Our
model suppose that \textit{there is no fluid }($\rho=0)$\textit{
outside the sphere of fluid (}$\rho_{\text{VACUUM}}$). Given that
the fluid simulates space-time continuum itself, there is nothing
($\rho=0)$ outside the sphere. Everything happens as if our sphere
was a ``Universe''. By introducing mathematical constant $\Lambda$
in (1) we are thus naturally led to a theory of Universe, i.e. a cosmological
interpretation: 
\begin{equation}
\text{\ }R(t)=R_{H}e^{H_{\Lambda}t}\text{ \ \ \ }\Leftrightarrow\text{\ \ \ \ }R(t)=\frac{c}{H_{\Lambda}}e^{H_{\Lambda}t}\text{ \ }\label{16}
\end{equation}
Our model explains then why Hubble's expansion is necessarily a global
expansion (no local observed effect of expansion. If the constant
of \textit{integration} $R_{H}$ (15-16), i.e. a \textit{global} constant,
is not equal to $c/H$ (if $\dot{R}(0)\neq c,$ see Remark 3), there
would exist two global constants of Hubble. This would be a nonsense.
In order to have Pseudo-Newtonian model of Pseudo-Euclidean fluid.
without contradiction, we must have $H_{\Lambda}R_{H}=c$ (16)%
\footnote{Our pseudo-Newtonian model, entirely based on (12), cannot be confused
with historical Friedman's Newtonian model. Friedman considered a
sphere of material fluid with variable radius $R(t)$ and material
density $\rho(t)$ but, unlike (12), with constant mass $M$ of Universe
$M=V(t)\rho(t)=\frac{4}{3}\pi R^{3}(t)\rho(t)\Rightarrow\frac{dM}{dt}=\frac{d}{dR}(\rho R^{3})=0$.
He considered a test body at radial distance with radial velocity
$\dot{R}(t)$ and therefore he obtained an equation with $E=0$ (in
standard notation $K=$ -$\frac{2E}{m}$ =0): $E=E_{c}+E_{p}=\frac{1}{2}m\dot{R}^{2}(t)-\frac{4}{3}\pi Gm\rho(t)R^{2}(t)=0\Rightarrow(\frac{\dot{R}(t)}{R(t)})^{2}=\frac{8\pi G\rho_{c}}{3}$.
With critical constant density he obtained $H(t)=H$ and therefore
(14). Nevertheless in Friedman's prerelativistic Newtonian model the
total mass $M$ of Universe is constant whilst, in our Pseudo-Newtonian
model, there is a typically relativistic variable mass of vacuum (c=1)
$M(t)$$=M_{H}e^{3H_{\Lambda}t}$with $M(0)=M_{H}=\frac{1}{2}\frac{c^{2}}{G}R_{H}=\frac{1}{2}\frac{c^{2}}{G}\sqrt{\frac{3}{\Lambda}}$.%
}. 

From $\frac{d}{dt}(\frac{\dot{R}(t)}{R(t)})=\frac{R(t)\ddot{R}(t)-\dot{R}(t)^{2}}{R^{2}(t)}=\dot{H}_{\Lambda}(t)=0$
we can define from radial acceleration $\ddot{R}(t)$ also a (standard)
parameter of deceleration that we suggest to note $q_{\Lambda}$:
\begin{equation}
q_{\Lambda}=-\frac{R(t)\ddot{R}(t)}{\dot{R}(t)^{2}}=-1\label{17}
\end{equation}
$q_{\Lambda}$ is here negative and implies thus an acceleration of
expansion. Initial condition of (17) $t=0$ are:
\begin{equation}
\frac{R(0)\ddot{R}(0)}{\dot{R}(0)^{2}}=\frac{R_{H}\alpha_{\Lambda}}{c^{2}}=1\tag{17-bis}
\end{equation}
Initial conditions mean that $R(0)=R_{H}$ and $\ddot{R}(0)=\alpha_{\Lambda}$
define ``horizon values'' exactly on the same way that $\dot{R}(0)=c$
defines an ``horizon value'' (Remark 3). We obtain a basic minimal
relativistic acceleration%
\footnote{We rediscover here in the framework of GR and SR, Milgrom's idea of
minimal acceleration. %
}. 

We deduce a pseudoNewtonian \textbf{scalar} field of gravitational
force with a global principle of equivalence ''acceleration-gravitation''
$\frac{\ddot{R}(t)}{2}=G\frac{M(t)}{R(t)^{2}}$. Our dynamical Universe
supposes then, with $\frac{1}{2}\dot{R^{2}}(0)-G\frac{M(0)}{R(0)}=0$,
an initial linear density $\frac{M(0)}{R(0)}=\frac{1}{2G}c^{2}$(together
with an initial force $\frac{1}{2G}c^{4}$ and power of expansion
$\frac{1}{2G}c^{5}$).

\textbf{\textit{REMARK 3 An important objection could be formulated
at this stage: Our Pseudo-Newtonian model would not be a Pseudo-Euclidean
model because our basic equation (12) uses a non-relativistic form
of energy}}\textit{. }

\textit{Everybody knows how to write kinetics energy for a material
particle $mc^{2}$$(\gamma(\beta)-1)$. Here we have not a material
point but a point of fluid in the framework of GR. Escape velocity
$\dot{R}(t)$ for such a point can be as large as we wish (not limited
by $c$). Pseudo-Newtonian equation (12) is a relativistic equation
because velocity of light plays a basic role. In fact, the escape
velocity $\dot{R}(t)$, of a point of space itself, is limited by
$c$ but not in usual meaning $\beta<c$ with domain of variation
$[0,c[$. Indeed, in order to avoid a supplementary constant $R_{H}$
(14)$,$ if we admit for the initial velocity (13) $\dot{R}(0)=cR_{H}\sqrt{\frac{\Lambda}{3}}=c$,
the domain $]c,\infty[$ of variation of $\dot{R}(t)$ is limited
by $c$. We rediscover in this way a basic tachyonic Pseudo-Euclidean
``light-space-time'' structure. We have to expect then optical properties
of fluid.}

\section{Optical properties of Fluid and Bondi's Doppler Redshift factor}

In Cosmology our model is very near the model of \textit{\emph{de
Sitter's empty (}}\emph{$\rho=p=0$}\textit{\emph{) Universe. The
latter is also in exponential expansion}}%
\footnote{In the model of Einstein-de Sitter (1932) there is also a Parabolic-Euclidean
solution ($K=0)$ with Critical density. Nevertheless it is a model
(with $\Lambda=0$) for usual matter. This model is today obsolete
because usual matter seems occupy only 1\% of critical density. \textit{\emph{Hoyle's
Steady State is also based on metric (18-de Sitter) with a null CC.
Minkowskian fluid involves a perfect cosmological principle (Hoyle-Bondi).}}%
}\emph{$a(t)=Ae^{\sqrt{\frac{\Lambda}{3}}t}$}\textit{\emph{ }}\emph{with
$H_{\Lambda}=\frac{\dot{a}(t)}{a(t)}$}\textit{\emph{\ and acceleration\ }}\emph{$q=-1$}\textit{\emph{.
Lemaitre' scale factor }}$a(t)$\textit{\emph{ is introduced in de
Sitter's metric (23)(\cite{5}).}}

\begin{equation}
ds^{2}=c^{2}dt^{2}-a^{2}(t)(dr^{2}+r^{2}d\theta^{2}+r^{2}\sin^{2}\theta d\phi^{2})
\end{equation}
\textit{\emph{whilst our scale factor }}\emph{$R(t)$}\textit{\emph{
is globally induced from (12)}}. In \textit{\emph{de Sitter's model,
the constant $A$ }}($"A=1"$)\textit{\emph{ in $a(t)=$$Ae^{\sqrt{\frac{\Lambda}{3}}t}$
is then not a global constant determined by initial condition of the
problem (}}like $R_{H}$\textit{\emph{ in 16). }}

\emph{With}\textit{\emph{ condition of radiality}}\emph{ $d\theta=d\phi=0$
}\textit{\emph{we have respectively}}\emph{ }\textit{\emph{de Sitter's
metric and Minkowski's metric (}}\emph{$a(t)=1$ with $R(t)\neq0$}

\emph{
\begin{equation}
ds^{2}=c^{2}dt^{2}-a^{2}(t)dr^{2}\,\;\; de\, Sitter\qquad ds^{2}=c^{2}dt^{2}-dr^{2}\,\; Minkowski\,\; CC\Rightarrow R(t)\label{21}
\end{equation}
}\textit{\emph{that are both particular cases of non-static (\cite{4})
radial Robertson-Walker's metric}}\emph{ $(dy=dz=0)$}
\begin{equation}
ds^{2}=c^{2}dt^{2}-\frac{a^{2}(t)}{1-Kr^{2}}dr^{2}
\end{equation}
\textit{\emph{with local (in metric) parameter of Gaussian curvature
}}\emph{$K=0$}. 

Let us now introduce the limit of light velocity with $ds^{2}=0$
first in flat metric of de Sitter $(K=0)$:

\[
dt^{2}-a^{2}(t)dr^{2}=0\qquad de\, Sitter\,(c=1)
\]
Usually one deduces from de Sitter model the following formula of
Redshift $z=\frac{\triangle\lambda}{\lambda}$ by Doppler effect in
$GR$ 
\begin{equation}
1+z=\frac{a(t_{reception})}{a(t_{emission})}
\end{equation}
(with standard notations of the time of emission of radial photon
from a remote galaxy towards time of reception in our galaxy). Moreover
with two usual cosmological measurable parameters $H$ and $q$ according
to the following standard development into series:
\begin{equation}
1+z\backsimeq1+Hr_{0}+(1+\frac{1}{2}q)H^{2}r_{0}^{2}+.=1+\beta_{s}+(1+\frac{1}{2}q)\beta_{s}^{2}+..
\end{equation}
where$\beta_{s}=Hr_{0}$ is standard law of Hubble (with radial comobile
distance$r_{0}$). Recall that$\beta_{s}$$=\frac{v}{c}$ (s for space)
is not the velocity between two galaxies (two material $\beta_{m}$
points) but a velocity $\beta_{s}=Hr_{0}$ between the ``points''
(elements of volume) of space itself occupied by galaxies. In de Sitter's
case we have thus:

\begin{equation}
1+z\backsimeq1+\beta_{s}+\frac{1}{2}\beta_{s}^{2}..
\end{equation}
Let us now introduce (inferior) velocity of light in our model. Optical
property of our fluid is given by Minkowskian limit ($dt^{2}-dr^{2}=0)$
of GR: \emph{
\begin{equation}
1+z=\frac{R(t_{reception})}{R(t_{emission})}=k_{s}=\sqrt{\frac{1+\beta_{s}}{1-\beta_{s}}}\label{21-1}
\end{equation}
}It is logical to admit a special relativistic development (Einstein's
standard Doppler radial factor for material point $k_{m}=\sqrt{\frac{1+\beta_{m}}{1-\beta_{m}}}$)
where $z$ can be as large that we wish$(\beta_{s}$$<1)$%
\footnote{We have therefore $\beta_{S}=\frac{v}{c}$ $v<c$ and $\beta_{S}=\frac{r_{0}}{R_{H}}$
$r_{0}<R_{H}$ as well ($\beta_{S}=\frac{v}{c}=Hr_{0}$). Everything
happens as if, underlying our model, there is a new DSR (Doubly Special
Relativity) a SR with a second constant $R_{H}$, i.e. a basic (maximal)
length $R_{H}$ (in Minkowskian scale hyperbolas) directly connected
with$\Lambda$(or a minimal acceleration).%
}. Equation (24) involves then a ``GR interpretation'' (velocity
of space itself) of Einstein's Doppler formula. With $q=-1$(in 25)
we have precisely (23) until the second order.

\begin{equation}
k_{s}=k_{Bondi}=\sqrt{\frac{1+\beta_{s}}{1-\beta_{s}}}\backsimeq1+\beta_{s}+\frac{1}{2}\beta_{s}^{2}+\frac{1}{2}\beta_{s}^{3}+\frac{3}{8}\beta_{s}^{4}...
\end{equation}
Our conjecture (24) here suggest then the conjecture that the parameters
of accelerating universe is given by famous ``Bondi's factor'' \textit{at
any order (25).} 

For the coherence of our model of points of space without baryonic
mass, we need for the light a null rest mass in such a way that in
perfect fluid (in 3), we have $"p+\rho=0"$ in front of the term of
four-velocity $\frac{u_{\mu}u_{\nu}}{c^{2}}$ of ``particle'' (see
note 2 Poincaré's electron)%
\footnote{Minkowskian limit can be also directly induced from equation of perfect
fluid (3). If we introduce Einstein's photon (light complex in 1905),
a particle of null rest mass (unlike Poincaré's purely wavy representation
of light), we have in this case in (3) $"p+\rho=0"$ or more precisely
$"p_{E}+\frac{1}{3}w_{em}=0"$. Invariant Minkowskian light cone implies
then a quantum definition of light with $p_{E}=$$-\frac{1}{3}w_{em}$.
This quantum definition of light is not historically a complete surprise
because it is already in Einstein's original parper (1905). Einstein's
transformation of Energy of light complex and Poincaré's transformation
of Energy of electron are together deduced with the same Lorentz transformation
of a moving sphere around a moving point. In both case, we have then
a strange pressure. Einstein considered a moving sphere at light velocity
with SPHERICAL density of PLANE wave ! SIC!. Lorentz (and Planck)
considered that it was impossible in the framework of classical em.
When he discover the famous (quantum) proportionality between energy
and frequency of light complex, the young statistical thermodynamician
had added non only particles of null rest mass but also a strange
(gravitational) pressure.

Let us note finally that in our scalar field, gravitational redshift
(that is usually determined in GR also with Einstein's photon!) is
then inseparable of usual redshift. %
}.

\section{Conclusion: Accelerating Flat Universe, Scalar Field and Dark Energy}

We showed the existence of a simple unexpected global Minkowskian
solution of Einstein's complete (with CC) equation of GR. The \textit{logical}
sequence from Pseudo-Euclidean solution (2) towards the Pseudo-Newtonian
Fluid (12) is the following $(c=1)$:

\[
T_{\mu\nu}^{VIDE}=\frac{\Lambda}{8\pi G}\mathbf{\eta}_{\mu\nu}\text{ \ }\Longrightarrow p+\rho=0\Longrightarrow dU(t)+pdV(t)=0\Longrightarrow\frac{1}{2}\dot{R}(t)^{2}-\frac{4}{3}\pi G\rho R(t)^{2}=0
\]
Minkowskian metric (infinitesimal interval) involves (with CC) then
a global scale factor $R(t)$. We wonder if we can introduce such
a scale factor in a finite interval in a next paper\cite{12}. From
relativistic pseudoNewtonian equation (12), we deduce here dynamical
properties $H_{\Lambda}=\frac{\dot{R}(t)}{R(t)}$, $q_{\Lambda}=-\frac{R(t)\ddot{R}(t)}{\dot{R}(t)^{2}}=-1$
and optical property (24) $1+z=\frac{R(t_{reception})}{R(t_{emission})}=k$
with Bondi's factor reinterpreted as a Redshift in $GR$. 

Dynamical (§1, §2, §3) and optical properties of our Minkowskian fluid
(or Continuum) are thus compatible with the most recent cosmological
observations (\cite{8}, \cite{9} \& \cite{10}):

1° Hubble's Redshift,

$2\lyxmathsym{°}$Parameter of curvature near zero ($K=0)$, 

3° Density near ''critical density '',

$4\lyxmathsym{°}$Parameter of acceleration near $q=-1$, 

5° Dark energy connected with a non-null $CC$ (note 6). 

Unlike usual Quantum approach of Vacuum (Lema\^{i}tre) our approach
consists in simulating properties of Vacuum with a ''Classical (apparently
at the departure) Continuum''. With quantum representation of light
(note 10), our model becomes compatible, for example, with a continuum
spectrum of a ``black body'' in Universal Vacuum.

\bigskip{}

\end{document}